\def\sL{\scriptscriptstyle L}
\def\sR{\scriptscriptstyle R}
\def\sY{\scriptscriptstyle Y}

\documentstyle[12pt]{article}

\begin{document}
\begin{flushright}
\begin{tabular}{l}
HUPD-9401\hspace{0.5cm}\\
February 1994
\end{tabular}
\end{flushright}
\vglue 0.9cm
\begin{center}
{{\Huge Dynamical CP Violation\\[0.3cm]
        in Composite Higgs Models\\[0.3cm]
        - A Review -\\}
\vglue 1.2cm
{\Large T.~Inagaki\hspace{-0.1cm}%
\renewcommand{\thefootnote}{}
\footnote{\normalsize
Talk presented at the Third KEK Topical Conference on CP Violation,
16-18 November, 1993.
The main part of this talk is based on the work in
collaboration with
S.~Hashimoto and T.~Muta [1].}\\}
\baselineskip=13pt
\vglue 0.8cm
{\it Department of Physics, Hiroshima University\\}
\baselineskip=12pt
\vglue 0.6cm
{\it Higashi-Hiroshima, Hiroshima 724, Japan\\}
\vglue 3.5cm

{\LARGE Abstract}}
\end{center}
\vglue 0.4cm
{\rightskip=3pc
 \leftskip=3pc
 \baselineskip=25pt
 \noindent

In composite Higgs models it was pointed out that there is a
possibility to violate CP symmetry dynamically.
We demonstrated a simple model of dynamical CP violation in
composite Higgs models.
We calculated the neutron electric dipole moment in
our model and the constraint for our model is discussed below.

}

\newpage

\baselineskip=25pt

\section{Introduction}

In the standard theory CP violating phenomena are described by phases
appearing in the Kobayashi-Maskawa matrix \cite{KM}.
The CP violating phases in KM matrix are introduced as free parameters.
This situation is not satisfactory for the fundamental theory of quarks and
leptons.
We would like to see what the origin of CP violation is.\\
\hspace*{\parindent}
One possibility to explain the origin of CP violation was pointed out by
T.~D.~Lee \cite{SCP}.
The idea is the following:
If the vacuum expectation value of Higgs fields is not real, the vacuum state
is not symmetric under CP transformation.
Thus CP symmetry is broken spontaneously.
This mechanism suggests that the spontaneous electroweak symmetry breaking
has something to do with the origin of CP violation.\\
\hspace*{\parindent}
On the other hand there is a possibility that the Higgs fields may be
constructed as bound states of more
fundamental fermions.
In this case the electroweak symmetry is broken down dynamically by the vacuum
expectation value of composite fields constructed by fermions and
anti-fermions.
Eichten, Lane and Preskill applied the idea of the spontaneous CP violation
to one of the composite Higgs models, the technicolor model \cite{ELP}.
They pointed out that through phases of the vacuum expectation value of the
composite Higgs field the CP symmetry is broken.
If the vacuum expectation value of composite fields has the complex phase, CP
symmetry is broken dynamically.
We call this mechanism the dynamical CP violation \cite{DCP}.
As a result of the dynamical CP violation in composite Higgs models CP
violating four-fermion interactions generally show up.
In this talk first I will briefly explain a mechanism of the dynamical CP
violation in composite Higgs models.
Then I will construct a simple model of the dynamical CP violation.
Finally I will discuss the neutron electric dipole moment in our model.

\section{Mechanism of Dynamical CP Violation}

In composite Higgs models the Higgs particle appears as a bound state of
fundamental fermion $Q$.
\begin{equation}
     \phi \sim \bar{Q}Q \, .
\end{equation}
There is a variety of composite Higgs models including the technicolor
model \cite{TC}, top-quark condensation model \cite{TOP},
top-color model \cite{TOPC}, fourth-generation model \cite{FG} and
color-sextet quark model \cite{SEX}.
The Lagrangian for the theory considered now is described as
\begin{equation}
     {\cal L}_{0} ={\cal L}_{QCD}+{\cal L}_{EW}+{\cal L}_{dyn}
\end{equation}
where ${\cal L}_{QCD}$ is the QCD Lagrangian without mass terms and
${\cal L}_{EW}$ is the electroweak Lagrangian without the elementary Higgs
fields and ${\cal L}_{dyn}$ is the Lagrangian describing the unknown dynamics
among fundamental fermions.
We assume that the Lagrangian has the global flavor symmetry.\\
\hspace*{\parindent}
To induce the electroweak symmetry breaking dynamically we require that the
vacuum expectation value of the composite fields is non-vanishing.
Thus we assume that ${\cal L}_{dyn}$ generates the fermion-antifermion
condensation.
As is well-known, ${\cal L}_{QCD}$ also generates the quark-antiquak
condensation.
\begin{equation}
\langle \bar{Q}Q \rangle \neq 0 \, , \mbox{\hspace{5mm}}
\langle \bar{q}q \rangle \neq 0 \, ,
\end{equation}
where $Q$ denotes the fundamental fermion and $q$ denotes the ordinary quark.
These fermion-antifermion condensations break the chiral symmetry
included in the global flavor symmetry.
Thus we get the theory with highly degenerate vacua.
Out of these degenerate vacua we can choose the vacuum for which
all vacuum expectation values of the composite fields is real.
Therefore the CP symmetry is not broken.\\
\hspace*{\parindent}
We need the flavor symmetry breaking terms ${\cal L}'$ to
introduce the mass hierarchy of ordinary quarks in the theory.
Of course in the case of standard model such terms are described by the Yukawa
interactions ${\cal L}_{\sY}$.
\begin{equation}
       {\cal L}_{\sY}=y^u \bar{q}_{\sL}u_{\sR}\phi
                   +y^d \bar{q}_{\sL}d_{\sR}\tilde{\phi}+h.c.
\end{equation}
If we replace the elementary scalar field $\phi$ by the composite field
$\bar{Q}Q$ we generate four-fermion interactions.
\begin{equation}
        {\cal L}_{4f}=G\bar{q}_{\sL}q_{\sR}\bar{Q}_{\sR}Q_{\sL}+h.c.
\end{equation}
These terms are the flavor symmetry breaking terms in the ordinary composite
Higgs models.
We suppose that these terms correspond to the low energy effective Lagrangian
stemming from the more fundamental Lagrangian at extremely high energy scale.\\
\hspace*{\parindent}
Let us see the vacuum state is symmetric or not under CP
transformation when these small flavor symmetry breaking terms are
added to the flavor symmetric Lagrangian ${\cal L}_{0}$.
We require the CP invariance of the full Lagrangian.
Then all the coupling $G$ of the four-fermion interactions should be real.
In this case the degeneracy of the ground state is solved by the flavor
symmetry breaking terms ${\cal L}'$ and the unique vacuum is determined.
We no longer have the degree of freedom to choose the vacuum for which the
vacuum expectation value of the composite fields is real.
Hence the vacuum expectation value of composite fields has a
complex phase.
This means that for the vacuum the CP symmetry is broken.
We need to make a transformation on the field under the global flavor symmetry
to make the vacuum expectation value real.
By this transformation the form of the flavor symmetry breaking terms
${\cal L}'$ are modified so that CP violating terms, in general, show up in
${\cal L}'$.
Thus the dynamical CP violation occurs.
In the following section I would like to present a simple model of the
dynamical CP violation in composite Higgs models.

\section{A simple model}

To see the characteristics of the dynamical CP violation in composite Higgs
models we construct a simple model here.
We assume the presence of two generations of extra-fundamental fermions
and three generations of ordinary quarks.
\begin{equation}
        Q\sim \left(
              \begin{array}{c}
                   U\\
                   D
              \end{array}
              \right)\, ,
              \left(
              \begin{array}{c}
                   C\\
                   S
              \end{array}
              \right)\, ,
\end{equation}
\begin{equation}
        q\sim \left(
              \begin{array}{c}
                   u\\
                   d
              \end{array}
              \right)\, ,
              \left(
              \begin{array}{c}
                   c\\
                   s
              \end{array}
              \right)\, ,
              \left(
              \begin{array}{c}
                   t\\
                   b
              \end{array}
              \right)\, .
\end{equation}
Our Lagrangian here is divided into two parts.
The full Lagrangian does not violate the CP symmetry.\\
\hspace*{\parindent}
${\cal L}_{0}$ is symmetric part under the global flavor symmetry $G_F$.
The flavor symmetry is
\begin{equation}
     G_{F}=U_{L}^{Q}(2)\otimes U_{R}^{Q}(2)
   \otimes U_{L}^{q}(3)\otimes U_{R}^{q}(3)\, .
\end{equation}
The part $U_{L}^{Q}(2)\otimes U_{R}^{Q}(2)$ is the flavor symmetry of
fundamental fermions and the other part $U_{L}^{q}(3)\otimes U_{R}^{q}(3)$ is
the flavor symmetry of ordinary quarks.
We consider only the up-type fermions for simplicity below.
Of course our model equally applies to the system of the down-type quarks.
The composite operators of fermion-antifermion develop the non-vanishing
vacuum expectation values under the Lagrangian ${\cal L}_{0}$.
\begin{equation}
\begin{array}{l}
     \langle \bar{U}_{\sL}U_{\sR} \rangle
   = \langle \bar{C}_{\sL}C_{\sR} \rangle \neq 0 \, ,\\
     \langle \bar{u}_{\sL}u_{\sR} \rangle
   = \langle \bar{c}_{\sL}c_{\sR} \rangle
   = \langle \bar{t}_{\sL}t_{\sR} \rangle \neq 0 \, .
\end{array}
\end{equation}
CP symmetry is not broken dynamically under the flavor
symmetric part ${\cal L}_{0}$.
We can take the vacuum for which all of these vacuum
expectation values are real.
The complex phases appear by the effect of the flavor symmetry
breaking part ${\cal L}'$.
By a transformation under the global flavor symmetry $G_F$,
arbitrary complex phases are generated in the vacuum expectation values.
\begin{equation}
     \begin{array}{ll}
          Q_{\sL}\longrightarrow W_{\sL}^{Q}Q_{\sL}\, , &
          Q_{\sR}\longrightarrow W_{\sR}^{Q}Q_{\sR}\, , \\
          q_{\sL}\longrightarrow W_{\sL}^{q}q_{\sL}\, , &
          q_{\sR}\longrightarrow W_{\sR}^{q}q_{\sR}\, ,
     \end{array}
\end{equation}
where $W_{\sL}^{Q}$, $W_{\sR}^{Q}$, $W_{\sL}^{q}$ and $W_{\sR}^{q}$ are
elements of flavor symmetries $U_{L}^{Q}(2)$, $U_{R}^{Q}(2)$,
$U_{L}^{q}(3)$ and $U_{R}^{q}(3)$.
We can cancel out the phases using this transformation.
However the flavor symmetry breaking part ${\cal L}'$ is not
symmetric under this transformation, then the form of ${\cal L}'$ will be
modified.\\
\hspace*{\parindent}
In composite Higgs models the flavor symmetry breaking part of the Lagrangian
is described by the four-fermion interactions.
\begin{eqnarray}
     {\cal L}' = && G_{1}^{Q}\bar{U}_{\sL}U_{\sR}\bar{U}_{\sR}U_{\sL}
                  + G_{2}^{Q}\bar{C}_{\sL}C_{\sR}\bar{C}_{\sR}C_{\sL}
                                                        \nonumber \\
                &+& (G_{1}^{Qq}\bar{U}_{\sL}U_{\sR}
                   + G_{2}^{Qq}\bar{C}_{\sL}C_{\sR})
                    (\bar{u}_{\sR}u_{\sL} + \bar{c}_{\sR}c_{\sL})
                                                        \nonumber \\
                &+& (G_{3}^{Qq}\bar{U}_{\sL}U_{\sR}
                   - G_{4}^{Qq}\bar{C}_{\sL}C_{\sR})
                     \bar{t}_{\sR}t_{\sL}
                                                        \nonumber \\
                &+& G^{q}\bar{q}_{\sL}q_{\sR}\bar{q}_{\sR}q_{\sL}\, .
\end{eqnarray}
Here we neglect the difference between up quark and charm quark.
As we require the CP invariance, all of the couplings $G$ have to be real.
If we take all of the couplings to be positive or all of the couplings to be
negative, the CP symmetry is not broken dynamically.
Hence we suppose that all of $G$ are positive and take a minus sign only for a
coupling $G_{4}^{Qq}$.
To find the true vacuum state of the full Lagrangian,
we start from one of the vacua
for the flavor symmetric part ${\cal L}_{0}$,
transform the vacuum by $W_L$ and $W_R$ to find $\langle -{\cal L}'\rangle$
under the transformed vacuum, and then try to minimize
$\langle -{\cal L}'\rangle$ by changing $W_L$ and $W_R$.
To calculate this energy $\langle -{\cal L}'\rangle$ it is convenient to
parametrize $W_L$ and $W_R$ in this way.
\begin{equation}
\begin{array}{l}
     (W_{\sR}^{Q}{W_{\sL}^{Q}}^{\scriptscriptstyle \dagger})
    = w_{i}^{Q}\exp (i\theta_{i}^{Q})\, , \\
     (W_{\sR}^{q}{W_{\sL}^{q}}^{\scriptscriptstyle \dagger})
    = w_{i}^{q}\exp (i\theta_{i}^{q})\, .
\end{array}
\end{equation}
Using this parametrization $\langle -{\cal L}'\rangle$ is described as a
function of $w$ and $\theta$.
\begin{eqnarray}
     \langle -{\cal L}' \rangle =
                  &-& (G_{1}^{Q}(w_{1}^{Q})^{2}
                     + G_{2}^{Q}(w_{2}^{Q})^{2})\Delta^{Q}
                                                        \nonumber \\
                &-& G_{1}^{Qq}(w_{1}^{Q}w_{1}^{q}
                               e^{i(\theta_{1}^{Q}-\theta_{1}^{q})}
                             + w_{1}^{Q}w_{2}^{q}
                               e^{i(\theta_{1}^{Q}-\theta_{2}^{q})})\Delta^{Qq}
                                                        \nonumber \\
                &-& G_{2}^{Qq}(w_{2}^{Q}w_{1}^{q}
                               e^{i(\theta_{2}^{Q}-\theta_{1}^{q})}
                             + w_{2}^{Q}w_{2}^{q}
                               e^{i(\theta_{2}^{Q}-\theta_{2}^{q})})\Delta^{Qq}
                                                        \nonumber \\
                &-& G_{3}^{Qq}w_{1}^{Q}w_{3}^{q}
                               e^{i(\theta_{1}^{Q}-\theta_{3}^{q})}\Delta^{Qq}
                                                        \nonumber \\
                &+& G_{4}^{Qq}w_{2}^{Q}w_{3}^{q}
                               e^{i(\theta_{2}^{Q}-\theta_{3}^{q})}\Delta^{Qq}
                                                        \nonumber \\
                &+& \mbox{O}(r^{2})\, .
\end{eqnarray}
In the vacua for ${\cal L}_{0}$ we may express the amplitude
\begin{eqnarray}
     \langle \bar{Q}_{\sL}Q_{\sR}\bar{Q}_{\sR}Q_{\sL} \rangle  =  \Delta^{Q}
                                                        \, ,\nonumber \\
     \langle \bar{Q}_{\sL}Q_{\sR}\bar{q}_{\sR}q_{\sL} \rangle  =  \Delta^{Qq}
                                                        \, ,\nonumber \\
     \langle \bar{q}_{\sL}q_{\sR}\bar{q}_{\sR}q_{\sL} \rangle  =  \Delta^{q}
                                                        \, .
\end{eqnarray}
We try to find which $w$ and $\theta$ minimize the energy
$\langle -{\cal L}'\rangle$.
To find the solution we assume
\begin{equation}
     r \sim \frac{\Delta^{Qq}}{\Delta^{Q}}
       \sim \frac{\Delta^{q}}{\Delta^{Qq}} \sim \mbox{O}(10^{-9})\, ,
\end{equation}
and in the following we neglect the terms of O$(r^2)$.\\
\hspace*{\parindent}
After some calculations we find that these relations have to be satisfied
to minimize the energy $\langle -{\cal L}'\rangle$.
\begin{equation}
      w_{i}^{Q}=w_{i}^{q}=1\, ,
\end{equation}
\begin{eqnarray}
           \frac{G_{1}^{Qq}}{G_{3}^{Qq}}
     & = & -2\frac{\sin (\theta_{1}^{Q}-\theta_{3}^{q})}
                  {\sin (\theta_{1}^{Q}-\theta_{1}^{q})}\, ,   \nonumber \\
           \frac{G_{2}^{Qq}}{G_{4}^{Qq}}
     & = & 2\frac{\sin (\theta_{1}^{Q}-\theta_{3}^{q})}
                  {\sin (\theta_{2}^{Q}-\theta_{1}^{q})}\, ,   \\
           \frac{G_{1}^{Qq}}{G_{2}^{Qq}}
     & = & -\frac{\sin (\theta_{2}^{Q}-\theta_{1}^{q})}
                  {\sin (\theta_{1}^{Q}-\theta_{1}^{q})}\, .   \nonumber
\end{eqnarray}
Non-vanishing $\theta$ means that the complex phase appears in the expectation
value of the composite field and the CP symmetry is broken.
So we would like to see if there are any couplings $G$ to satisfy these
relations for non-vanishing $\theta$.\\
\hspace*{\parindent}
To find a realistic solution we assume that the strong CP violation
should be absent for ordinary quarks.
This means that
\begin{equation}
     {\theta}_{1}^{q}+{\theta}_{2}^{q}+{\theta}_{3}^{q}
    =2{\theta}_{1}^{q}+{\theta}_{3}^{q}=0\, .
\label{eqn:scp}
\end{equation}
Here we neglect the difference between the up quark and the charm quark and
we take
${\theta}_1={\theta}_2$
We take into account that the top quark mass is $100$ times heavier than the
up quark mass and the charm quark mass.
This means that the four-fermion couplings relevant to the top quark is larger
than those for the up and the charm quark.
\begin{equation}
\begin{array}{l}
     \displaystyle \frac{G_{1}^{Qq}}{G_{3}^{Qq}} \sim \mbox{O}(1/10)\, ,\\
     \displaystyle \frac{G_{2}^{Qq}}{G_{4}^{Qq}} \sim \mbox{O}(1/10)\, ,\\
     \displaystyle \frac{G_{1}^{Qq}}{G_{2}^{Qq}} \sim \mbox{O}(1)\, .
\end{array}
\end{equation}
\hspace*{\parindent}
Here we notice that there are three equations and four parameters to be
determined so that we
have the degree of freedom to satisfy the relation (\ref{eqn:scp}).
It should be noted that we have only one degree of freedom and so we can
apply this relation to the ordinary quark sector only.
Thus the strong CP problem is not resolved in the fundamental fermion sector.\\
\hspace*{\parindent}
After some calculation we get the solution which satisfies these relations.
\begin{equation}
\begin{array}{l}
     \displaystyle \theta_{1}^{Q}=-\epsilon_{1}+\frac{2}{3}\theta\, , \\
     \displaystyle \theta_{2}^{Q}=\pi+\epsilon_{2}+\frac{2}{3}\theta\, , \\
     \displaystyle \theta_{1}^{q}=\theta_{2}^{q}=-\frac{1}{3}\theta\, ,\\
     \displaystyle \theta_{3}^{q}=\frac{2}{3}\theta\, .
\end{array}
\end{equation}
Parameters ${\epsilon}_1$, ${\epsilon}_2$ and $\theta$ are defined by
\begin{equation}
\begin{array}{l}
     \displaystyle
     \frac{G_{1}^{Qq}}{G_{3}^{Qq}}=\frac{\epsilon_{1}}{\sin \theta}\, ,
                                                                   \\
     \displaystyle
     \frac{G_{2}^{Qq}}{G_{4}^{Qq}}=\frac{\epsilon_{2}}{\sin \theta}\, ,
                                                                   \\
     \displaystyle
     \frac{G_{1}^{Qq}}{G_{2}^{Qq}}
    =1+(\epsilon_{1}+\epsilon_{2})\frac{\cos \theta}{\sin \theta}\, .
\end{array}
\end{equation}
The condition that the top quark is 100 times heavier than the up and the
charm quark lead to
\begin{equation}
     \epsilon_{1} \sim \epsilon_{2} \sim \mbox{O}(1/10)\, ,\mbox{\hspace{5mm}}
     \sin \theta \sim \mbox{O}(1)\, .
\label{eqn:sin}
\end{equation}
\hspace*{\parindent}
And this solution shows that
\begin{equation}
     W_{R}^{Q} {W_{L}^{Q}}^{\scriptscriptstyle {\dagger}}=
     \left(
     \begin{array}{cc}
     e^{i(-\epsilon_{1}+\frac{2}{3}\theta)} &    \\
     & e^{i(\pi+\epsilon_{2}+\frac{2}{3}\theta)}
     \end{array}
     \right)\, ,
\end{equation}
\begin{equation}
     W_{R}^{q} {W_{L}^{q}}^{\scriptscriptstyle {\dagger}}=
     \left(
     \begin{array}{ccc}
     e^{-i\frac{1}{3}\theta} &  &  \\
     & e^{-i\frac{1}{3}\theta} &   \\
     &  & e^{i\frac{2}{3}\theta}
     \end{array}
     \right)\, .
\end{equation}
These phases correspond to the phases of the vacuum expectation value of the
composite operators.
Accordingly the CP symmetry is broken.\\
\hspace*{\parindent}
We make inverse-transformations to obtain the real vacuum expectation value
of the composite fields.
\begin{equation}
     \begin{array}{ll}
          Q_{\sL}\longrightarrow {W_{\sL}^{Q}}^{\scriptscriptstyle \dagger}
                                  Q_{\sL}\, , &
          Q_{\sR}\longrightarrow {W_{\sR}^{Q}}^{\scriptscriptstyle \dagger}
                                  Q_{\sL}\, , \\
          q_{\sL}\longrightarrow {W_{\sL}^{q}}^{\scriptscriptstyle \dagger}
                                  q_{\sL}\, , &
          q_{\sR}\longrightarrow {W_{\sR}^{q}}^{\scriptscriptstyle \dagger}
                                  q_{\sL}\, .
     \end{array}
\end{equation}
The form of the four-fermion terms ${\cal L}'$ is modified and we get the CP
violating four-fermion interactions.
\begin{eqnarray}
     {\cal L}' = && G_{1}^{Q}\bar{U}_{\sL}U_{\sR}\bar{U}_{\sR}U_{\sL}
                  + G_{2}^{Q}\bar{C}_{\sL}C_{\sR}\bar{C}_{\sR}C_{\sL}
                                                        \nonumber \\
                &+& (G_{1}^{Qq}e^{i(\theta-\epsilon_{1})}
                     \bar{U}_{\sL}U_{\sR}
                   + G_{2}^{Qq}e^{i(\pi+\theta+\epsilon_{2})}
                     \bar{C}_{\sL}C_{\sR})
                    (\bar{u}_{\sR}u_{\sL} + \bar{c}_{\sR}c_{\sL})
                                                        \nonumber \\
                &+& (G_{3}^{Qq}e^{i\epsilon_{1}}\bar{U}_{\sL}U_{\sR}
                   - G_{4}^{Qq}e^{i(\pi+\epsilon_{2})}\bar{C}_{\sL}C_{\sR})
                    \bar{t}_{\sR}t_{\sL}
                                                        \nonumber \\
                &+& G^{q}_{1111}\bar{u}_{\sL}u_{\sR}\bar{u}_{\sR}u_{\sL}
                   + \cdots                             \nonumber \\
                &+& G^{q}_{1313}e^{i\theta}
                    \bar{u}_{\sL}t_{\sR}\bar{u}_{\sR}t_{\sL}
                   + \cdots \, .
\end{eqnarray}
\hspace*{\parindent}
As we neglect the difference between the up quark and the charm quark,
no relative phase appears between up and charm.
And the parts describing the four-fermion interaction between ordinary
quarks are important for the low energy phenomena.\\
\hspace*{\parindent}
We confirm that the dynamical CP violation occurs in our simple model.
Our model is too simple to explain the Kobayashi-Maskawa matrix and should
be elaborated to reproduce the standard theory as  a low energy effective
theory.
If our model has something to do with nature, it has to be consistent with
the existing experimental observations.
For this purpose we calculate the neutron electric dipole moment under our
Lagrangian.

\section{Neutron electric dipole moment in our model}

It is well-known that the neutron electric dipole moment calculated in the
standard theory is extremely small \cite{NEDM}.
Thus it is possible to observe an extra effect to the neutron
electric dipole moment coming from the CP violating four-fermion interactions.
Here we calculate the neutron electric dipole moment from the CP violating
four-fermion Lagrangian of ordinary quarks obtained in the last section.
\begin{equation}
     {\cal L}' = {G_{ijkl}}'
                    \bar{q}_{i\sL}q_{j\sR}\bar{q}_{k\sR}q_{l\sL}\, ,
\end{equation}
where indices $i,j,k,l$ represent flavor of q, i.e., $u, d, c, s, t, b.$
and coupling constants $G'$ generally have complex phases.
\hspace*{\parindent}
The neutron electric dipole moment $d_n$ is given in terms of the quark dipole
moments $d_u$ and $d_d$ in the naive quark model such that
\begin{equation}
     d_{n}=\frac{1}{3}(4d_{d}-d_{u})
\end{equation}
The upper bound for the neutron electric dipole moment $d_n$ imposed
by the experimental observation is $d_{n}<10^{-25} \mbox{e cm}.$
The electric dipole moment of quarks is calculated through the
following term in the quark electromagnetic form factor at
zero-momentum transfer
\begin{equation}
     -d_{q}\bar{u}_{q}\sigma_{\mu\nu}\gamma_{5}q^{\nu}u_{q}
\end{equation}
where $q^{\nu}$ is the momentum transfer for quarks (momentum carried by the
virtual photon) and $u$ is the Dirac spinor for quark $q$.\\
\hspace*{\parindent}
At one loop-level, the diagrams shown in Fig.1 contribute to the
quark electromagnetic form factor.
However these diagrams do not contribute to the electric dipole moment.
At two loop level,  we show the  diagrams  which contribute to the
quark electromagnetic form factor in Fig.2.
Because of the quadratically divergent parts of the amplitude,
the diagrams shown in Fig. 2(c) have the largest contribution to the quark
electric dipole moment.

\begin{figure}
\vspace{70pt}
\caption{One-loop diagrams for the electromagnetic vertex function of quarks.}
\label{fig:oneloop}
\end{figure}

\begin{figure}
\vspace{70pt}
\caption{Two-loop diagrams for the electromagnetic vertex function of quarks.}
\label{fig:twoloop}
\end{figure}

\hspace*{\parindent}
We calculate the quadratically divergent parts of the amplitude and
find the main contribution to the quark electric dipole moment.
\begin{equation}
     d_{u} =  \frac{2}{3}\mbox{e}\frac{\Lambda^{2}}{(4\pi)^{4}}\sum_{i,j,k}
           \mbox{Im}({G_{jiuk}}'{G_{kjiu}}')m_{j}
           \left( 2\mbox{ln}\frac{\Lambda^{2}}{m_{j}^{2}}
                              -2.01 \right) \, ,
\end{equation}
where $\Lambda$ is the cut-off parameter.
Here we assume
\begin{equation}
     |{G_{jiuk}}'{G_{kjiu}}'| \sim \frac{g^4}{4\Lambda^{4}}
                \sim \frac{4\pi^2}{4\Lambda^{4}} \, ,
\end{equation}
where $g$ is the coupling constant for the fundamental theory at high
energy scale $\Lambda$ and we assume
\begin{equation}
     \frac{g^2}{4\pi} \sim 1 \, .
\end{equation}
And the top quark mass is much heavier than the other quark masses.
Only the term proportional to the top quark mass has a large contribution.
Thus we find
\begin{equation}
     d_{u} \sim \mbox{e}\frac{m_{t}}{48\pi^{2}\Lambda^{2}}\sin \theta
                \left( 2\mbox{ln}\frac{\Lambda^{2}}{m_{t}^{2}}
                                  -2.01 \right)\, .
\end{equation}
If we suppose that the top quark mass is about $140$ GeV and use the upper
bound of the neutron electric dipole moment $d_n$ from the experiments
and the constraint shown Eq.(\ref{eqn:sin}) we
obtain the lower bound of the cut-off scale $\Lambda$.
\begin{equation}
     \Lambda > 800 \mbox{TeV}\, .
\end{equation}
This lower bound of the cut-off scale $\Lambda$ is of the same order as the
one set by the flavor changing neutral current in kaon physics \cite{FCNC}.

\section{CONCLUSION}

We succeeded in finding a simple model of dynamical CP violation.
And to satisfy the experimental constraint from neutron electric dipole
moment in our model the cut-off
scale of the theory has to be higher than $800$ GeV.\\
\hspace*{\parindent}
There are some remaining problems.
To explain the actual KM matrix our simple model has to be further elaborated.
The strong CP problem in the sector of fundamental fermions is still an open
problem.
We are aware of the cosmological domain wall problem with regard to the
dynamical breaking of the discrete symmetry and leave it to the future
research.

\section*{Acknowledgement}

I would like to thank T.~Muta for valuable discussions
and T.~Kouno for inserting figures in this manuscript.
I am also grateful to G.~Burnett for kindly reading this manuscript.


\begin{thebibliography}{99}
\bibitem{HIM} S.~Hashimoto, T.~Inagaki and T.~Muta,
              {\it Phys. Rev.} {\bf D48}, (1993) 1301.
\bibitem{KM}  M.~Kobayashi and T.~Maskawa, {\it Prog. Theor. Phys.} {\bf 49}
                                                               , (1973) 652.
\bibitem{SCP} S.~Weinberg, {\it Phys. Rev. Lett.} {\bf 37}, (1976) 657.\\
              T.~D.~Lee, {\it Phys. Rev.} {\bf D8}, (1973) 1226;
                                  {\it Phys. Reports} {\bf 9C}, (1974) 143.
\bibitem{ELP} E.~Eichten, K.~Lane and J.~Preskill, {\it Phys. Rev. Lett.}
                                                    {\bf 45}, (1980) 225.
\bibitem{DCP} R.~Dashen, {\it Phys. Rev.} {\bf D3}, (1971) 1879.\\
              W.~Goldstein, {\it Nucl. Phys.} {\bf B213}, (1983) 477;
                                                     {\bf B229}, (1983) 157.
\bibitem{TC} S.~Weinberg, {\it Phys. Rev.} {\bf D13}, (1976) 974;
                            {\bf D19}, (1978) 1277.\\
               L.~Susskind, {\it ibid.} {\bf D20}, (1979) 2619.
\bibitem{TOP} Y.~Nambu,
              {\it Proc. Intern. Workshop on New Trends in Strong Coupling
               Gauge Theories}, Nagoya, 1988,
               eds. M.~Bando, T.~Muta and K.~Yamawaki
               (World Scientific Pub. Co., 1989).\\
               V.~A.~Miransky, M.~Tanabashi and K.~Yamawaki,
               {\it Phys. Lett.} {\bf B211}, (1989) 177;
               {\it Mod. Phys. Lett.} {\bf A4}, (1989) 1043.\\
               W.~J.~Marciano, {\it Phys. Rev. Lett.} {\bf 62}, (1989) 2793;
               {\it Phys. Rev.} {\bf D41}, (1990) 219.\\
               W.~A.~Bardeen, C.~T.~Hill and M.~Lindner, {\it ibid.}
               {\bf D41}, (1990) 1647.
\bibitem{TOPC} C.~T.~Hill, {\it Phis. Lett.} {\bf B266}, (1991) 419.
\bibitem{FG} C.~T.~Hill, M.~A.~Luty and E.~A.~Paschos, {\it Phys. Rev.}
             {\bf D43}, (1991) 3011.
\bibitem{SEX} W.~J.~Marciano, {\it Phys. Rev.} {\bf D21}, (1980) 2425.\\
              K.~Fukazawa, T.~Muta, J.~Saito, I.~Wata-\\
              nabe, M.~Yonezawa and M.~Inoue,
              {\it Prog. Theor. Phys.} {\bf 85}, (1991) 111.
\bibitem{NEDM} X.-I.~He, B.~H.~J.~McKellar and S.~Pakvasa,
               {\it Int. J. Mod. Phys.}
               {\bf A4}, (1989) 5011.
\bibitem{FCNC} R.~N.~Cahn and H.~Harari,
                               {\it Nucl. Phys.} {\bf B176}, (1980) 135.\\
               S.~Dimopoulos and J.~Ellis,
                               {\it ibid.} {\bf B182}, (1981) 505.

\end{thebibliography}
\end{document}